\documentclass[aps,prl,amsmath,amsfonts,showpacs]{revtex4}

\begin{document}

\title{More about the stringy limit of black hole equilibria}

\author{Bernard~S~Kay\footnote{Electronic address: bernard.kay@york.ac.uk}}

\affiliation{Department of Mathematics, University of York, York YO10 5DD, U.K.}

\begin{abstract}
We discuss further our proposed modification of the Susskind-Horowitz-Polchinski scenario in which black hole entropy goes over to string entropy as one scales the string length scale up and the string coupling constant down, keeping Newton's constant unchanged.  In our approach, based on our {\it matter-gravity entanglement hypothesis}, `string entropy' here should be interpreted as the likely entanglement entropy between (approximately) the single long string and the stringy atmosphere which, as we argue, arise in a pure multistring state of given energy in a (rescaled) box.  In a previous simple analysis, we computed this entropy (with promising results) by assuming simple exponentially increasing densities of states for both long string and stringy atmosphere.  Here, our starting point is a (more correct) density of states for each single string with the appropriate inverse-power prefactor and a low-energy cutoff.  We outline how the relevant entanglement entropy should be calculated for this system and propose a plausible model, which draws on the work of Mitchell and Turok on the multi-string microcanonical ensemble, in which we adopt a similar density of states for long string and stringy atmosphere but now with cutoffs which scale with the total energy, $E$.  With this scaling, we still find our entanglement entropy grows, in leading order, linearly with $E$ and this translates to a black-hole entropy which goes as the square of the black-hole mass, thus retaining most of the promising features of our previous exponential-density-of-states model and providing further evidence for our matter-gravity entanglement hypothesis.
\end{abstract}


\maketitle

In \cite{Kay1, Kay2, KayAbyaneh} we proposed the \textit{matter-gravity entanglement hypothesis} according to which the physical entropy of a general (quantum gravitational) closed system should be identified with its \textit{matter-gravity entanglement entropy}.  In particular \cite {Kay1, Kay2, KayAbyaneh, Kaythermality}, a black hole equilibrium state, consisting of a (say spherical, 4-dimensional) black hole in equilibrium with its (mostly [see Endnote (iii) in \cite{KayAby}]) matter atmosphere in a box, is assumed to be pure -- described by a total state vector, $\Psi$, of quantum gravity, of  given approximate energy, $E$, which is entangled in just such a way that the reduced density operator for gravity alone, and the reduced density operator for matter alone are both (approximately) thermal (i.e.\ Gibbs) states at the Hawking temperature.   In \cite{Kaythermality, Kaystringy}, we proposed (as a modification of the scenario proposed in \cite{Susskind} and \cite{HorowitzPolchinski}) that, in a string theory description, the weak string-coupling limit of such a black hole equilibrium state, obtained by scaling the string length scale, $\ell$, up and the string coupling constant, $g$, down from their physical values, keeping  $G=g^2\ell^2$ and $E$ fixed  ($G$ denotes Newton's constant and we set $c=\hbar=1$), will consist of a pure total state vector, which we shall also call $\Psi$, representing a long string in equilibrium with an atmosphere of small strings in a suitably rescaled box.   We assumed that it is approximately \cite{gravitons} correct to make the identifications: long string $\leftrightarrow$ gravity and stringy atmosphere $\leftrightarrow$ matter and, in particular, that the (matter-gravity entanglement) entropy of the black hole equilibrium state will be approximately equal to the  the long string-stringy atmosphere entanglement entropy of the rescaled $\Psi$, provided, when (with a modification of the, semi-qualitative, procedure of \cite{Susskind, HorowitzPolchinski} suited to our matter-gravity entanglement hypothesis)  we make the identification when 
\begin{equation}
\label{ellXGM}
\ell=XG{\cal M}
\end{equation}
where $X$ is a numerical factor of order 1.  And we assume that $\cal M$ can be equated with the mean energy of the long string.  To calculate this entropy in this way, we assumed, in \cite{Kaythermality, Kaystringy}, a simplified model, according to which both long string and stringy atmosphere are quantum systems with exponentially growing densities of states, 
\begin{equation}
\label{expdenstates}
\sigma_{\mathrm{ls}}=C_{\mathrm{ls}}e^{\ell\epsilon}, \quad \sigma_{\mathrm{sa}}=C_{\mathrm{sa}}e^{\ell\epsilon}.
\end{equation}
 
In \cite{Kaythermality}, we showed that, quite generally, such an approximate energy eigenstate of a total quantum system consisting of two weakly coupled subsystems denoted `S' and `B' \cite{terminology} -- with given densities of states, $\sigma_{\mathrm{S}}$ and $\sigma_{\mathrm{B}}$, will have thermodynamic properties which are largely independent of the details of the total state vector, $\Psi$, other than that it has a given approximate energy.  More precisely, we defined the state space, ${\cal H}_M$, spanned by total energy eigenstates with energies in the range $[E, E+\Delta]$ where $\Delta$ is small, but large enough for the dimension, $M$,  of ${\cal H}_M$ to be large.  We then showed that, for a wide range of (positively supported, monotonically increasing) $\sigma_{\mathrm{S}}$ and $\sigma_{\mathrm{B}}$, if one chooses $\Psi$ at random from ${\cal H}_M$, then the mean energy (and other moments of energy) of each of S and B as well as the S-B entanglement entropy, $S$ (necessarily equal to the von Neumann entropy of S and also to the von Neumann entropy of B)  will, with high probability, be close to the values calculated from the `modapprox' density operators $\rho^{\mathrm{modapprox}}_{\mathrm{S}}$ and $\rho^{\mathrm{modapprox}}_{\mathrm{B}}$ (see \cite{Kaythermality} or \cite{Kaystringy} for the explicit formulae).  We show in \cite{Kaythermality} that, as a result of this, the (thus largely state-independent) S-B entanglement entropy will then be given by the formula \cite{terminology}
\begin{equation}
\label{contpureentropy}
S=
k\int_0^{E_c} P_{\mathrm{S}}(\epsilon)\log\left (
\frac{\sigma_{\mathrm{S}}(\epsilon)}{P_{\mathrm{S}}(\epsilon)}\right)d\epsilon+k\int_{E_c}^E P_{\mathrm{S}}(\epsilon)\log\left (
\frac{\sigma_{\mathrm{B}}(E-\epsilon)}{P_{\mathrm{S}}(\epsilon)}\right)d\epsilon
\end{equation}
where $E_c$ is the energy at which $\sigma_{\mathrm{S}}(E_c)=\sigma_{\mathrm{B}}(E-E_c)$, $P_{\mathrm{S}}(\epsilon)$ is the {\it energy-probability density} of S
\begin{equation}
\label{enprobdens}
P_{\mathrm{S}}(\epsilon)=
\frac{\Delta}{M}\sigma_{\mathrm{S}}(\epsilon)\sigma_{\mathrm{B}}(E-\epsilon)
\end{equation}
and $k$ denotes Boltzmann's constant.

Applying (\ref{enprobdens}) to the densities of states (\ref{expdenstates}), the mean energy, $\bar\epsilon_{\mathrm{ls}}$, of the long string ($=\int_0^\infty \epsilon P_{\mathrm{B}}(\epsilon) d\epsilon) =  E/2$ and, similarly (or, in consequence) the mean energy, $\bar\epsilon_{\mathrm{sa}}$, of the stringy atmosphere $=E/2$ and, as we showed in \cite{Kaythermality}, applying (\ref{contpureentropy}) we find $S=k\ell E/4$.   Equating $\bar\epsilon_{\mathrm{ls}}$ with the black hole mass, $\cal M$, this model thus predicts, by (\ref{ellXGM}), that the entropy of the black hole equilibrium state is $kXG{\cal M}^2/2$.  Moreover, we showed in  \cite{Kaythermality} that, for densities of states as in (\ref{expdenstates}), the modapprox density operators of S and B (i.e.\ of the long string and the stringy atmosphere) are each (in a precise sense made clear there) {\it approximately thermal} with inverse temperature $k\ell$ which, by (\ref{ellXGM}) is $kXG{\cal M}$.  As we remarked in \cite{Kaythermality, Kaystringy} it is an intriguing coincidence that, by taking $X$ to have the (single) value $8\pi$, these results simultaneously agree with the Hawking entropy formula, $S=4\pi G{\cal M}^2$ and with the inverse Hawking temperature formula, $S=8\pi G{\cal M}$, for a Schwarzschild black hole. 

The above, exactly `exponential', model is inadequate in a number of respects:  No reason is given as to why the string equilibrium state should consist of a single long string and an atmosphere of small strings and (related to this) the exponential form assumed for the densities of states of the long string and of the string atmosphere is too crude an approximation.  Also, the result seems, in a number of respects, to be physically unrealistic:  By (\ref{enprobdens}), the energy probability density of the long string takes the flat (see Figure 3 in \cite{Kaythermality}) form  $P_{\mathrm{ls}}(\epsilon)=1/E$ (and similarly for the string atmosphere)
which entails unrealistic-seeming large fluctuations in the distribution of energy between the long string and the stringy atmosphere.  Moreover it seems unnatural that the mean energy of the long string (and hence also of the string atmosphere) is predicted to be exactly $E/2$.  Presumably, it should really be a fraction of $E$ which depends on the size of the box.

In the present letter, we report on an attempt to overcome these limitations with a more sophisticated model.  Our starting point is a system of many weakly-coupled (closed, bosonic) strings in a box of volume $V$ in $D+2$ spacetime dimensions (we shall mainly be interested in the case $D=2$ \cite{dimension}) -- each with the physically more realistic single-string density of states 
\begin{equation}
\label{singstringdens}
\sigma_{\mathrm{ss}}(\epsilon)=f(\epsilon)e^{\ell\epsilon}
\end{equation}
where, for $\epsilon$ greater than some cutoff, $\epsilon_0$, the pre-factor, $f(\epsilon)$, takes the power-law form
\begin{equation}
\label{powerlaw}
f(\epsilon) = C_{\mathrm{ss}} \epsilon^{-p}
\end{equation}
where $p=(D+3)/2$ \cite{inverse power} and $C_{\mathrm{ss}}$ is a constant of order $\ell^{-(D+1)/2}$, while $f(\epsilon)=0$ for $\epsilon$ below a cutoff, $\epsilon_0$.   This cutoff is needed because, for energies below the gap, $\sqrt{2D/3}(\pi/\ell)$, between the lowest two string levels, the prefactor (\ref{powerlaw}) spuriously increases as the energy, $\epsilon$, decreases, diverging at $\epsilon=0$; (\ref{powerlaw}) also significantly overestimates the correct density of states for energies up to a few times this `gap' value.  We expect to achieve a best fit to the results which would be obtained for the true density of states by choosing $\epsilon_0$ close to the value of this gap (and, in any case, $O(1/\ell)$).

Mitchell and Turok \cite{MitchellTurok} have studied the microcanonical ensemble for such a system of weakly coupled strings with a total-momentum zero constraint and with a cutoff, $m_0$, somewhat larger than that we envisage here for $\epsilon_0$, in fact, large enough for a certain non-relativistic approximation, which they make, to be valid.    They find that, for total energies, $E$, and total volumes, $V$, such that the energy density, $\rho=E/V$, is very much greater than the natural string energy density, $\ell^{-(D+2)}$, the multi-string configurations in the energy range $[E, E+\Delta]$ will consist predominantly of configurations where there is one long string with an atmosphere of small strings in which the probability of a string having a mass, $m$, is (provided it is bigger than $m_0$) proportional to $m^{-(D+3)/2}$.  As they point out, this corresponds to a `scale-invariant' distribution of string lengths and, more relevantly to us here, is also consistent with the string atmosphere being thermal with inverse temperature $k\ell$.  They also find that the total number, $n$, of strings in the stringy atmosphere is close to a value, $n_{\mathrm{max}}$, of the order of $V$ divided by the natural string volume, $\ell^3$ (times a factor which becomes $O(1)$ if $m_0$ is replaced by our [order $1/\ell$] $\epsilon_0$ -- see below).   The above scale invariant mass distribution together with the cutoff at $m_0$ easily implies that the mean energy of the string atmosphere will be $((D+1)/(D-1))n_{\mathrm{max}}m_0$ and hence that the mean energy of the long string will be $E$ minus this.   (On the other hand, if $\rho=E/V$ is less than $\ell^{-(D+2)}$, they find a distribution of small loops in which large loops are exponentially suppressed.)

We shall assume that all these conclusions of \cite{MitchellTurok} will continue to hold when the cutoff, $m_0$, of \cite{MitchellTurok} is replaced by our (somewhat lower, $O(1/\ell)$) cutoff, $\epsilon_0$, even though the non-relativistic approximation of \cite{MitchellTurok} will then cease to be accurate and we shall also assume that the regime with one long string and a scale-invariant distribution of small strings will continue to be hold provide only $E$ is only a few times (rather than very much greater than) $V\ell^{-(D+2)}$, i.e., by the above quoted results and the above assumptions, moderately greater than a constant of $O(1)$ times $n_{\mathrm{max}}\epsilon_0$.  Thus, comparing the above result on the energy of the string atmosphere, the long string and the string atmosphere can have energies which are comparable in size.  But while, conceivably, the energy of the long string could be equal to or even slightly less than the energy of the stringy atmosphere, it cannot be much smaller, while the energy of the string atmosphere can easily be much smaller than the energy of the long string.

It seems plausible that, under the rescaling we discussed at the outset, this microcanonical string scenario will go over to a microcanonical scenario for quantum gravity in the strong-string-coupling regime, in which, in the high energy density regime, the long string goes over to a black hole and the stringy atmosphere to the black hole's atmosphere, while, in the low energy density regime, it goes over to a state involving matter particles and gravitons (see Endnote \cite{gravitons}) but no black holes.  (We are unaware of such a connection between string and black-hole microcanonical ensembles having been made previously in the literature.)   However, we are interested, not in a microcanonical scenario, but rather,  in the language of \cite{Kaythermality, Kaystringy}, in a {\it modern} scenario consistent with our matter-gravity entanglement hypothesis.  Nevertheless, the above results are relevant because the configurations which dominate the microcanonical ensemble at given $E$ and $V$ are, of course, dominant amongst the the state-vectors in the relevant  ${\cal H}_M$.    We propose the following {\it ab initio} approach to calculating the black hole entropy on this modern scenario: First, one realizes the Hilbert space, ${\cal H}_M$, of multistring states in our volume $V$ and with energies in an interval $[E, E+\Delta]$ (approximately) as a subspace of a tensor product, ${\cal H}_{\mathrm{sa}}\otimes {\cal H}_{\mathrm{ls}}$, of a {\it stringy atmosphere} Hilbert space, ${\cal H}_{\mathrm{sa}}$ (isomorphic to the Hilbert space of all multistring states) with a {\it long-string} Hilbert space, ${\cal H}_{\mathrm{ls}}$ (isomorphic to the Hilbert space of a single string).  To do this, one writes a general multistring state, with, say, $n$ strings, as $|s_1, \dots, s_n\rangle$ where $s_1, \dots, s_n$ denote the states of the individual strings, {\it listed with the highest-mass string -- i.e.\ the longest-length string --  first}.  This must of course be symmetrized appropriately for bosonic strings.   Then we define the map, $\mu$ by 
\begin{equation}
\label{longstringmap}
\mu: |s_1, s_2, \dots s_n\rangle \mapsto |s_2,\dots,s_n\rangle \otimes |s_1\rangle .
\end{equation} 
$\mu$ won't be exactly unitary (onto its range) because of the loss of full symmetrization.  However, we expect, from the results of \cite{MitchellTurok}, that, if $E/V$ is very much greater than the natural string energy density, then, very nearly always, the most massive (i.e.\ longest-length) string state, $|s_1\rangle$, will be far more massive (far longer) than any of the other string states $|s_2\rangle$, $\dots$ and hence very nearly always very nearly orthogonal to all of them.  Thus $\mu$ will be very close to unitary.   Proceeding as if it were exactly unitary, we can then use $\mu$ to identify ${\cal H}_M$ with the Hilbert space consisting of the range of $\mu$ in ${\cal H}_{\mathrm{sa}}\otimes {\cal H}_{\mathrm{ls}}$.  We then expect to find (cf. the results in \cite{Kaythermality}) that most states in this Hilbert space have a (long-string--string atmosphere) entanglement entropy close to a particular value.  It remains to verify this and to calculate that value. 

There is something necessarily partly vague about this:  As we anticipated in earlier work on our Matter Gravity Entanglement Hypothesis (see e.g.\ Endnote (ii) in \cite{KayAbyaneh}) the split in our degrees of freedom between `gravity' and `matter' as well as the notion of `black hole' all arise as emergent features of our underlying string theory.   As $E/V$ decreases towards, and then goes below, the natural string energy density, we will find that $|s_1\rangle$ has a mass/length less and less often distinctly larger than the typical mass/length of the other strings ($|s_2\rangle$, $\dots$), $\mu$ will be less and less accurately unitary and the `entanglement entropy' we calculate as above will (after the scaling procedure outlined at the outset) have less and less to do with matter-gravity entanglement entropy.  Furthermore, to calculate the matter-gravity entanglement entropy for moderately large $E/V$, we also expect to need to take into account gravitons \cite{gravitons} and, in the low $E/V$ regime, we expect it  almost entirely to arise as matter-graviton entanglement entropy.  Such entropies will, in the absence of `black holes', no doubt, be much lower but, as anticipated in our preliminary calculations in \cite{Kay2}, not totally zero.  We further remark that all this gives further grounds for taking seriously the suggestion, which we arrived at on the basis of independent arguments in Section IXA of \cite{Kaythermality}, that the density of states of matter (if not also of gravity) in full quantum gravity (which will, indirectly, be connected to that of our string theory through Equation (\ref{ellXGM})) should be `state-dependent'; in fact, in the present context, the suggestion is that it will be dependent on the total energy, $E$ and the volume (which will be a suitable rescaling of $V$).

We will not attempt such an {\it ab initio} approach here.  Rather we wish to propose what seems to be a reasonable guess for a model which, if correct, permits us to estimate the above entanglement entropy, in the large $E/V$ regime, by an application of the formalism of \cite{Kaythermality, Kaystringy}:  We replace the range of $\mu$ by a tensor product, say ${\tilde{\cal H}}_{\mathrm{sa}}\otimes {\tilde {\cal H}}_{\mathrm{ls}}$, and assume there is a total `effective Hamiltonian' on this Hilbert space which arises as the sum of a {\it string-atmosphere} Hamiltonian, $H_{\mathrm{sa}}\otimes I$, with a {\it long-string} Hamiltonian, $I\otimes H_{\mathrm{ls}}$, where the associated densities of states, $\sigma_{\mathrm{sa}}$ and $\sigma_{\mathrm{ls}}$,
both behave as (\ref{singstringdens}) with the power-law pre-factors as in (\ref{powerlaw}), both with the same power, $p=(D+3)/2$, as for a single string but possibly different constants, say $\tilde C_{\mathrm{sa}}$ and $\tilde C_{\mathrm{ls}}$.  But, to treat the case with total energy, $E$, we replace the cutoff $\epsilon_0$ with the cutoffs:  
\begin{equation}
\label{cutoffs}
\epsilon_{\mathrm{sa}}=\beta E, \quad \epsilon_{\mathrm{ls}}=\alpha E
\end{equation}
where $\alpha+\beta < 1$ and, importantly, $\alpha$  and $\beta$ are independent of $E$ (but will depend on $V$ in a way consistent with the way $n_{\mathrm{max}}$ scales with $V$).
The reason for the $\epsilon_{sa}$ cutoff is that we assume the total number of strings in the stringy atmosphere to be close to the  $n_{\mathrm{max}}$ of \cite{MitchellTurok}.  Then clearly, since each of these strings has an energy bounded below by $\epsilon_0$, the sum of all their energies will be bounded below by  $n_{\mathrm{max}}\epsilon_0$ ($=$ a constant of $O(1)$ times $V\ell^{-(D+2)}$) which as we saw above is a constant of $O(1)$ times $E$.  The reason for the $\epsilon_{ls}$ cutoff is that we require the energy of the long string to be comparable in size to $E$ or we would get a contradiction with the conditions for the existence of a long string at all!   We also know that $\beta$ may be rather small, while it becomes problematic if $\beta$ becomes close to or slightly larger than $\alpha$.

Strictly of course, in view of these $E$-dependent cutoffs, $\sigma_{\mathrm{sa}}$ and $\sigma_{\mathrm{ls}}$ are not ordinary densities of states; rather (assuming this model to be sound) the fact that they need to be cut off in this way provides further evidence (cf. above and Section IXA in \cite{Kaythermality}) that, in quantum gravity, one in a sense has state-dependent densities of states.  In spite of them being state-dependent however, we can still clearly apply the formalism of \cite{Kaythermality, Kaystringy}.

Amongst the weaknesses of the model are that we cannot say much more about the precise values of $\alpha$ and $\beta$ (although one hopes that they would effectively be determined in terms of $E$ and $V$ in an {\it ab initio} approach and in particular be confirmed to be independent of $E$).  Also the model does not explicitly include the constraint that the total momentum is zero, although some of the consequences of this condition are effectively incorporated, indirectly, in our cutoffs.

Amongst the arguments in favor of the model, first we note that the microcanonical ensemble appropriate to our densities of states, $\sigma_{\mathrm{sa}}$ and $\sigma_{\mathrm{ls}}$, will easily imply (e.g.\ by the methods in \cite{Kaythermality}) that the stringy atmosphere will be approximately thermal at inverse temperature, $k\ell$, in (approximate) agreement with the result of \cite{MitchellTurok} at least in the non-relativistic limit where the energy of each string is approximately its mass.   Secondly, we have the (at least approximate) consistency result: If we assume the above density of states, $\sigma_{\mathrm{sa}}$, for a system of $n$ strings (we indicate the dependence on $n$ with a superscript) then, for a system of $n+1$ strings, we will have $\sigma_{\mathrm{sa}}^{n+1}(\epsilon)=\int_0^\epsilon\sigma_{\mathrm{ss}}(\epsilon')\sigma_{sa}^n(\epsilon-\epsilon') d\epsilon'$
=$C_{\mathrm{ss}}C_\mathrm{sa}^n\int_{\epsilon_0}^{1-n\epsilon_0}\epsilon'^{-p}e^{\ell\epsilon}(\epsilon-\epsilon')^{-p}e^{\ell(\epsilon-\epsilon')}d\epsilon'$.  Simplifying and writing $\epsilon'=x\epsilon$ this is
$C_{\mathrm{ss}}C_\mathrm{sa}^n \epsilon^{-2p+1}e^{\ell\epsilon}\int_{\epsilon_0/\epsilon}^{1-n\epsilon_0/\epsilon}x^{-p}(1-x)^{-p}dx$.  Clearly, (for large $n$ and) for $\epsilon \gg \epsilon_0$, the latter integral is dominated by the behaviour at the lower limit, with approximate value $(\epsilon_0/\epsilon)^{-p+1}/(p-1)$.  Hence, for $\epsilon\gg \epsilon_0$,
$\sigma_{\mathrm{sa}}^{n+1}(\epsilon)\simeq C_{\mathrm{sa}}^{n+1}\epsilon^{-p}e^{\ell\epsilon}$ where
$C_{\mathrm{sa}}^{n+1}\simeq C_{\mathrm{ss}}C_\mathrm{sa}^n\epsilon_0^{-p+1}/(p-1)$ (above, $p=(D+3)/2$).  And, obviously, $\sigma_{\mathrm{sa}}^{n+1}=0$ for  $\epsilon < (n+1)\epsilon_0$.

Thirdly, if we assume the string atmosphere to consist exactly of $n_{\mathrm{max}}$ strings then, when $E$ is very much greater than (or, indeed, just a few multiples of)  $n_{\mathrm{max}}\epsilon_0$, the joint energy distribution, $\sigma_{\mathrm{ss}}(\epsilon_1)
\dots\sigma_{\mathrm{ss}}(\epsilon_n)$, for fixed total energy, $\epsilon_1 + \dots +\epsilon_n = E$ will (by (\ref{singstringdens}) with (\ref{powerlaw}) and taking logarithms) clearly be maximized when all but one of the strings have energy close to $\epsilon_0$ while the remaining string has to have enough energy to bring the total up to $E$, and therefore (at least in the non-relativistic regime) must be `long', in accord with the results of \cite{MitchellTurok}.  (But the full analysis of \cite{MitchellTurok} is still needed to show that $n$ is peaked around an $n_{\mathrm{max}}$ which is independent of $E$ [in fact a constant of $O(1)$ times $V\ell^{-(D+1)}$]). 

With this model, we easily have that the stringy atmosphere's energy probability density (cf. (\ref{enprobdens})), $P_{\mathrm{sa}}(\epsilon)={\cal N} \epsilon^{-p}(E-\epsilon)^{-p}$ for $\epsilon\in[\beta E, (1-\alpha)E]$ and equals zero outside this interval, where ${\cal N}$ ensures $P_{\mathrm{sa}}$ integrates to 1. (And $P_{\mathrm{ls}}(\epsilon)=P_{\mathrm{sa}}(E-\epsilon)$.)  In fact, clearly, we have ${\cal N}=C^{-1}E^{2p-1}$ where $C=\int_\beta^{1-\alpha}x^{-p}(1-x)^{-p}dx$. Replacing the logarithms which appear in (\ref{contpureentropy}) by their leading terms, $\ell\epsilon$ and $\ell(E-\epsilon)$ respectively, we see that the entanglement entropy, $S$, between string atmosphere and long string is most probably very close to
${\cal N}k\ell\left(\int_{\beta E}^{E_c}\epsilon^{-p+1}(E-\epsilon)^pd\epsilon + \int_{E_c}^{(1-\alpha)E}\epsilon^{-p}(E-\epsilon)^{-p+1}d\epsilon\right)$
plus subleading terms involving logarithms.  From now on ignoring the latter, and temporarily specializing to cases where $\alpha$ and $\beta$ are both less than $1/2$ and $\tilde C_{\mathrm{sa}}=\tilde C_\mathrm{ls}$  whereupon $E_c$ obviously simply takes the value $1/2$, we then find that 
\begin{equation}
\label{D}
S=kD\ell E \ \  \hbox{where} \ \   D=\frac{\int_\beta^{1/2} x^{-p+1}(1-x)^{-p}dx + \int_{1/2}^\alpha x^{-p}(1-x)^{-p+1}dx}{\int_\beta^\alpha
x^{-p}(1-x)^{-p}dx}.
\end{equation}
If we relax the condition $\tilde C_{\mathrm{sa}}=\tilde C_\mathrm{ls}$, $E_c$ will neither be a constant nor will it exactly scale with $E$ and so we will not be able to cleanly extract a ratio of $E$-independent integrals as above.  However, $S$ will still take the form $kD\ell E$, with $D$ of order 1, plus subleading terms \cite{2longstrings}.

Turning from `entropy' to `temperature':  In many cases (e.g.\ if $\tilde C_{\mathrm{sa}}=\tilde C_\mathrm{ls}$, as long as $\alpha < 1/2$)  $E_c$ in (\ref{contpureentropy}) will lie somewhere in-between $\beta$ and $1-\alpha$ and there will be senses (albeit, depending on the exact values of $\alpha$ and $\beta$, more or less weak than the notion of
`$E$-approximately semithermal' in \cite{Kaythermality}) in which both the stringy atmosphere and the long string are `approximately thermal' at inverse temperature $k\ell$.   We note that it could happen (e.g.\ it clearly would happen if  $\tilde C_{\mathrm{sa}}=\tilde C_\mathrm{ls}$ and $\alpha > 1/2$ [whereupon $\beta < 1/2$]) that $E_c= 1-\alpha$ and then the second integral in  (\ref{contpureentropy}) will vanish and $S$ will equal the microcanonical entropy (see \cite{Kaythermality, Kaystringy}) of the stringy atmosphere.  In this case, while the sense in which the stringy atmosphere is approximately thermal will become more strong, the sense in which the long string is thermal will become considerably weaker.   (The parameter range $\alpha< 1/2$ and $\beta > 1/2$ seems unrealistic.)   

In any case, to the extent to which the long string and stringy atmosphere can each be said to have an `(inverse) temperature' at all, it will be $k\ell$.  By (\ref{ellXGM}), our model thus predicts a black-hole inverse temperature of $kXG{\cal M}$.  For a Schwarzschild black hole, this requires $X=8\pi$.

If we further tentatively equate ${\cal M}$ with the mean value, $\int_0^E \epsilon P_{ls}(\epsilon)d\epsilon ={\cal N}\int_{\beta E}^{1-\alpha E}\epsilon^{-p}(E-\epsilon)^{-p+1}d\epsilon=C^{-1}\left(\int_\beta^{1-\alpha} x^{-p}(1-x)^{-p+1}dx\right)E$, of the long-string energy, we will clearly have, from (\ref{ellXGM}) and (\ref{D}), that the entropy of the black hole equilibrium will take the form $kYG{\cal M}^2/2$ for a $Y$ of order 1 and this will agree with the Hawking formula, $S=4k\pi G{\cal M}^2$,  if $Y=8\pi$ ($=X$).   This would then confirm the intriguing feature noted above for the simple `exponential' model of \cite{Kaythermality, Kaystringy} and would, in fact, amount to an explanation of the Hawking value, $1/4$, in the Bekenstein-Hawking entropy formula, $S=k(\mathrm{Area})/4$.  One can try to adjust $\alpha$ and $\beta$ to make this happen, but one would want to know if the resulting values are justified by an {\it ab initio} calculation.  It may anyway be, though,  that to ask for a precise agreeement of $Y$ and $X$  is asking more of the, after all, semi-qualitative, approach of \cite{Susskind, HorowitzPolchinski} (as adapted in \cite{Kaythermality, Kaystringy} and here) than we have  a right to expect it to provide.  It does remain important, though, to investigate whether an ab initio calculation does indeed reproduce the main features of our model calculation.  

Assuming it does, our model confirms the prediction, of the `exponential' model of \cite{Kaythermality, Kaystringy}, of an entanglement entropy in a (pure) total vector-state, $\Psi$, between long string and stringy atmosphere which (whatever the values of the uncertain parameters, $\alpha$ and $\beta$) goes linearly with total energy and hence, with (\ref{ellXGM}), of an entanglement entropy between black hole and (mostly) matter atmosphere which goes as the square of the black hole mass thus (bearing in mind Endnote \cite{gravitons}) adding further support  to our matter-gravity entanglement hypothesis.  (And it goes against the often-expressed opinion that the thermal atmosphere of a black hole cannot be of any major importance in accounting for black hole entropy.) 

Finally, our scenario obviously cannot work for extremal black holes since they have no atmosphere!  However,  in the brick-wall model of \cite{tHooftBrick, MukohyamaIsrael} the entropy of an arbitrary non-extremal black hole is given, with no loss of accuracy as one approaches extremality, by the (von Neumann) entropy of its thermal atmosphere.   As noted in \cite{KayAbyaneh, KayOrtiz}, the brick wall result seems consistent with our entanglement picture of black hole equilibria.   This all suggests that the Hawking entropy of an extremal black hole should be interpreted physically as the extremal limit of the entropy of a sequence of non-extremal black holes, whilst an extremal black hole itself is an unphysical idealization.


\begin{thebibliography}{99}

\bibitem{Kay1} B.S.~Kay, \textit{Entropy defined, entropy increase and decoherence understood,
and some black-hole puzzles solved} [arXiv:hep-th/9802172]

\bibitem{Kay2} B.S.~Kay, Class.\ Quant.\ Grav.\ {\bf 15}, L89 (1998)
[arXiv:hep-th/9810077]

\bibitem{KayAbyaneh} B.S.~Kay and V.~Abyaneh, \textit{Expectation values, experimental predictions,
events and entropy in quantum gravitationally decohered quantum
mechanics} [arXiv:0710.0992]

\bibitem{KayAby}
See particularly Endnote (iii) in \cite{KayAbyaneh}

\bibitem{Kaythermality} B.S.~Kay, \textit{On the origin of thermality} [to appear on arXiv simultaneously with this paper]

\bibitem{Kaystringy} B.S.~Kay, \textit{Modern foundations for thermodynamics and the stringy limit of black hole equilibria} [to appear on arXiv simultaneously with this paper]

\bibitem{gravitons} In the absence of a black hole, one expects a correspondence at low enough energies:  matter $\leftrightarrow$ low energy string states with spin less than or equal to 1 and gravity $\leftrightarrow$ lowest energy string states with spin 2 (gravitons).  When a black hole is present, a part of the thermal atmosphere of a black hole will still consist of gravitons but, arguably, only a small part.   Cf.  \cite{tHooftBrick, MukohyamaIsrael}.  We assume that the major part of the gravitational degrees of freedom becomes (under the rescaling described in the text) the long string and not too big an error will result from neglecting the gravitons.

\bibitem{Susskind} L.~Susskind,  {\it Some speculations about black hole entropy in string theory}, arXiv:hep-th/9309145

\bibitem{HorowitzPolchinski} G.~Horowitz and J.~Polchinski, Phys.\ Rev.\ D {\bf 55}, 6189 (1997),  G.~Horowitz, {\it Quantum states of black holes}.  In R.M.~Wald (editor) {\it Black Holes and Relativistic Stars}
(University of Chicago Press, Chicago 1998) [arXiv:gr-qc/9704072]

\bibitem{terminology}  In \cite{Kaythermality} and \cite{Kaystringy} we use the expression `totem'  instead of `total system' and the (Roman) symbols `S' and `B' we retain here for its two subsystems stand for `System' and `(Energy) Bath'.  In all three papers (i.e.\ \cite{Kaythermality}, \cite{Kaystringy} and the present paper) we reserve the symbol (italic) `$S$' to denote an entropy.  What we denote simply `$S$' in Equation (\ref{contpureentropy}) here is denoted `$S_{\mathrm{S}}$' in \cite{Kaythermality, Kaystringy} but of course, as we explain in \cite{Kaythermality, Kaystringy}, this is equal to $S_{\mathrm{B}}$ as well as being equal to the S-B entanglement entropy.

\bibitem{dimension}  Of course one ought not, in string theory, choose any other than the appropriate critical dimension and also we should include fermions with superstrings etc.   But it seems reasonable to hope that our main  conclusions will survive when strings are treated more properly.  In this respect, we follow the spirit of the work of \cite{HorowitzPolchinski} and we also note that, as explained in \cite{MitchellTurok}, one can get sensible answers for the thermodynamics of strings for general values of $D$, particularly $D=2$.

\bibitem{inverse power} The power $p=(D+3)/2$ applies if we don't impose the constraint that the centre of mass momentum is zero.  If we do impose this, the correct power is $D+2$.  See e.g.\ \cite{MitchellTurok} and also
K.~Huang and S.~Weinberg, Phys.\ Rev.\ Lett.\  {\bf 25}, 895 (1970).

\bibitem{MitchellTurok} D.~Mitchell and N.~Turok, Nucl.\ Phys.\ B {\bf 294}, 1138 (1987) and Phys.\ Rev.\ Lett.\  {\bf 58}, 1577 (1987).

\bibitem{2longstrings} If we were to ask about the likely entanglement entropy of a pair of (say, non-identical) strings in a pure state, $\Psi$, of energy around $E$, the calculation would be very similar to the above except that the limits to the relevant integrals over $E$ would now be at $\epsilon_0$ and $E-\epsilon_0$, and we would find a result of $O(1)$ in $E$.  What makes our string atmosphere-long string entanglement entropy of $O(E)$ is the fact that our cutoffs scale with $E$.  All this is in contrast to what happens \cite{Kaythermality, Kaystringy} if we assume exactly exponential densities of states both for the long string/string atmosphere system and for such a pair of strings; obviously such assumptions will give the same entropy ($S=k\ell E/4$) for both cases.  So, in this sense, the simple exponential model of \cite{Kaythermality, Kaystringy} was misleading.

\bibitem{tHooftBrick} G. 't Hooft, Nucl.\ Phys. {\bf B256}, 727 (1985)

\bibitem{MukohyamaIsrael} S.~Mukohyama and W.~Israel, Phys.\ Rev. D {\bf 58} 104005 (1998)

\bibitem{KayOrtiz} B.S.~Kay and L.~Ortiz, {\it Brick Walls and AdS/CFT}, arXiv:1111.6429

\end{thebibliography}
\end{document}